\documentclass[twocolumn,aps,superscriptaddress,showpacs,nofootinbib,floatfix]{revtex4}
\usepackage{listings}
\usepackage[utf8]{inputenc}
\usepackage{hyperref}
\usepackage{amsmath}
\usepackage{amssymb}
\usepackage{amsfonts}
\usepackage{tabularx}
\usepackage{booktabs}
\usepackage{graphicx}
\usepackage{color}
\usepackage{multirow}
\usepackage[inline]{enumitem}
\usepackage[T2A,T1]{fontenc}

\graphicspath{{fig/}}
\definecolor{theblue}{RGB}{0,50,230}

\usepackage{hyperref}
\hypersetup{
  colorlinks=true,
  linkcolor=theblue,
  citecolor=theblue,
  urlcolor=theblue
}

\newcommand {\avg}[1]{\ensuremath{\langle\kern-1.0pt\langle#1\rangle\kern-1.0pt\rangle}}

\newlength\cmsFigWidth

\usepackage[normalem]{ulem}  
\usepackage{soul,xcolor}

\renewcommand\sout{\bgroup \color{red} \ULdepth=-.5ex \ULset}

\begin{document}

\title{Multiplicity Scaling of Light Nuclei Production in Relativistic Heavy-Ion Collisions}

\author{Wenbin~Zhao}
\affiliation{Key Laboratory of Quark and Lepton Physics (MOE) \& Institute of Particle Physics,Central China Normal University, Wuhan 430079, China}
\author{Kai-jia Sun}
\affiliation{Cyclotron Institute and Department of Physics and Astronomy, Texas A\&M University, College Station, Texas 77843, USA}
\author{Che Ming Ko}
\affiliation{Cyclotron Institute and Department of Physics and Astronomy, Texas A\&M University, College Station, Texas 77843, USA}
\author{Xiaofeng Luo}
\affiliation{Key Laboratory of Quark and Lepton Physics (MOE) \& Institute of Particle Physics,Central China Normal University, Wuhan 430079, China}
\date{\today}

\begin{abstract}
Using the nucleon coalescence model based on kinetic freeze-out nucleons from the 3D MUSIC+UrQMD and the 2D VISHNU hybrid model with a crossover equation of state, we study the multiplicity dependence of deuteron  ($d$)  and triton  ($t$) production from central to peripheral Au+Au collisions at $\sqrt{s_\mathrm{NN}}=$ 7.7, 14.5, 19.6, 27, 39, 62.4 and 200 GeV and Pb+Pb at $\sqrt{s_{\rm NN}}=2.76$ TeV, respectively. It is found that the ratio $N_t  N_p/N_d^2$ of the proton yield $N_p$, deuteron yield $N_d$ and triton yield $N_t$ exhibits a scaling behavior in its multiplicity dependence, i.e., decreasing monotonically with increasing charged-particle multiplicity. A similar multiplicity scaling of this ratio is also found in the nucleon coalescence calculation based on kinetic freeze-out nucleons from a multiphase transport (AMPT) model. The scaling behavior of $N_t  N_p/N_d^2$ can be naturally explained by  the interplay between the sizes of light nuclei and the nucleon emission source.  We further argue that the multiplicity scaling of $N_t  N_p/N_d^2$ can be used to validate the production mechanism of light nuclei, and the collision energy dependence of this yield ratio can further serve as a baseline in the search for the QCD critical point in relativistic heavy-ion collisions.  
\end{abstract}

\maketitle
\section{Introduction}

Light nuclei production in high energy heavy-ion collisions have been extensively
studied both experimentally and theoretically~\cite{Ambrosini:1997bf,Armstrong:2000gz,Adler:2001uy,Adler:2004uy,Arsene:2010px,Abelev:2010rv,Agakishiev:2011ib,Yu:2017bxv,Adam:2015yta,Adam:2015vda,Anticic:2016ckv,Acharya:2017dmc,Chen:2018tnh,Braun-Munzinger:2018hat,Donigus:2020ctf}.
Besides its intrinsic interest, studying light nuclei production provides
the possibility to probe the local baryon density and the space-time structure
of the emission source in relativistic heavy-ion collisions~\cite{Csernai:1986qf,Scheibl:1998tk,Bellini:2018epz}.
However, the production mechanism for light nuclei in heavy-ion collisions
is still under debate. On the one hand, the statistical model, which assumes
that light nuclei, like hadrons, are produced from a thermally and chemically
equilibrated source, provides a good description of measured yields of
light (anti-)nuclei in central Pb+Pb collisions at the Large Hadron Collider
(LHC)~\cite{Andronic:2017pug,Vovchenko:2018fiy,Vovchenko:2020dmv}. On the
other hand, the coalescence model, which assumes that light nuclei are
formed from the recombination of kinetically freeze-out protons and neutrons~\cite{Mattiello:1995xg,Mattiello:1996gq,Chen:2003qj,Chen:2003ava},
can also successfully describe the transverse momentum spectra and the
collective flow of light nuclei in heavy-ion collisions~\cite{Shah:2015oha,Sun:2015jta,Sun:2015ulc,Botvina:2017yqz,Yin:2017qhg,Zhao:2018lyf,Zhao:2020irc,Kozhevnikova:2020bdb}.
In between these two extreme scenarios for light nuclei production in heavy-ion
collisions is the transport model, which aims to study how light nuclei
evolve during the hadronic evolution by including their production and
annihilation, such as the processes
$\pi +p+n \leftrightarrow d+\pi $ for the deuteron~\cite{Oh:2009gx,Oliinychenko:2018ugs}.
In this approach, it is assumed that the deuteron can exist in hot dense
hadronic matter, although its temperature of more than 100 MeV is significantly
higher than the deuteron binding energy of 2.2 MeV.

Recently, more experimental data on light nuclei production in relativistic
heavy-ion collisions have become available. For example, the STAR Collaboration
at Relativistic Heavy-Ion Collider (RHIC) and the ALICE Collaboration at
the LHC have collected a wealth of data on light nuclei, such as (anti-)deuteron
($\bar{d}$, $d$), (anti-)triton ($\bar{t}$, $t$) and (anti-)helium-3 ($^{3}
\bar{H}e$, $^{3}He$), in Au+Au collisions from 7.7 GeV to 200 GeV~\cite{Adamczyk:2016gfs,Adam:2019wnb,Zhang:2020ewj}
and Pb+Pb collisions at 2.76 TeV~\cite{Adam:2015vda,Acharya:2020lus,Acharya:2019ttn},
respectively. A noteworthy result from these experiments is the observation
that the yield ratio of triton, deuteron and proton,
$N_{t} N_{p}/N_{d}^{2}$, in central Au+Au collisions shows a non-monotonic
energy dependence with a peak around $\sqrt{s_{\mathrm{NN}}}=$20 GeV~\cite{Adam:2019wnb,Zhang:2020ewj}.
A physically interesting explanation for this non-monotonic behavior is
that it is due to the density fluctuations indeuced by the change in the quark-gluon plasma (QGP) to hadronic
matter phase transition from a crossover at small baryon chemical potential
to a first-order transition at large baryon chemical potential and the
associated critical point~\cite{Sun:2017xrx,Sun:2018jhg,Yu:2018kvh,Liu:2019nii,Sun:2020uoj,Deng:2020zxo,Luo:2017faz}.
Another interesting observation from these experiments is the suppressed
production of light nuclei in peripheral Au+Au collisions or in collisions
of small systems. A possible explanation for this suppression is due to
the non-negligible sizes of light nuclei compared to the size of the nucleon
emission source in such collisions~\cite{Scheibl:1998tk,Sun:2018mqq}. Alternatively, introducing
the canonical suppression in the thermal model can also lead to the suppression
of light nuclei production in collisions of small systems~\cite{Vovchenko:2018fiy,Vovchenko:2020dmv}.

In this paper, we investigate the multiplicity dependence of deuteron and
triton production from peripheral to central Au+Au collisions at
$\sqrt{s_{\mathrm{NN}}}$ = 7.7 - 200 GeV from the Beam Energy Scan (BES) program
at RHIC by using the nucleon coalescence model with the needed phase-space
distribution of nucleons generated by the 3D MUSIC+UrQMD hybrid model~\cite{Shen:2020jwv}
and also the AMPT model~\cite{Lin:2004en}. We further use the 2D VISHNU
hybrid model~\cite{Zhao:2017yhj} to study deuteron and triton production
in Pb+Pb collisions at 2.76 TeV. Specifically, nucleons at the kinetic
freeze-out of the expanding hadronic matter in heavy-ion collisions are
obtained from these models with a smooth crossover transition from the
QGP or the partonic matter to the hadronic matter without the presence
of any density fluctuations. Since the MUSIC+UrQMD hybrid model used in
the present study has been shown to nicely reproduce the measured yield
and collective flow of various hadrons in heavy-ion collisions at both
RHIC and LHC energies~\cite{Shen:2017bsr,Shen:2017fnn,Denicol:2018wdp,Shen:2018pty,Shen:2020jwv,Zhao:2017yhj,Zhao:2017rgg,Zhao:2020pty}, 
although with a different initial state, it provides a robust baseline
for the study of light nuclei production in the RHIC BES program. We find
from our study that the yield ratio $N_{t}N_{p}/N^{2}_{d}$ in collisions
from $\sqrt{s_{\mathrm{NN}}}$=7.7 GeV to 2.76 TeV is essentially independent
of the collision energy and decreases with increasing charged-particle
multiplicity. Such a multiplicity scaling of this ratio contradicts to
the prediction of the thermal model, which shows an opposite multiplicity
dependence, i.e., an increasing with increasing charged-particle multiplicity~\cite{Vovchenko:2018fiy,Vovchenko:2020dmv}.

This paper is organized as follows: Section~\ref{sec:model} briefly introduces
the nucleon coalescence model for deuteron and triton production, and the
MUSIC+UrQMD hybrid model for the description of heavy- ion collision dynamics
at RHIC. Section~\ref{sec:results} presents and discusses results from
the MUSIC+UrQMD hybrid model on the multiplicity dependence of the transverse
momentum spectra and the rapidity distributions of protons, deuterons and
tritons, the coalescence parameters for deuteron and triton production,
and the yield ratio $N_{t} N_{p}/N_{d}^{2}$ in Au+Au collisions at RHIC
BES energies. The obtained charged particle multiplicity dependence of
the yield ratio $N_{t} N_{p}/N_{d}^{2}$ is then compared with that from
the AMPT+coalescence model and also extended to Pb+Pb collisions at LHC
by using the VISHNU+coalescence model. Section~\ref{sec:summary} concludes
this paper with a summary.

\section{The theoretical framework}\label{sec:model}

\subsection{The coalescence model for light nuclei production}\label{coalescence}

In the coalescence model~\cite{Mattiello:1995xg,Mattiello:1996gq,Chen:2003qj,Chen:2003ava},   the number of  a  light  nucleus of atomic number $A$ and consisting of $Z$ protons and $N$ neutrons ($A=Z+N$) is given by the overlap of the Wigner function  $f_A$ of the nucleus with the phase-space distributions $f_{p}({\bf x}_i, {\bf p}_i, t)$ of protons and  $f_{n}({\bf x}_j, {\bf p}_j, t)$ of neutrons~\cite{Chen:2003qj,Chen:2003ava},  
\begin{eqnarray}
\label{coal}
&&\frac{dN_A}{d^3 {\mathbf P}_A}= g_A \int \Pi_{i=1}^Zp_i^\mu d^3\sigma_{i\mu}\frac{d^3{\bf p}_i}{E_i}f_{p/\bar{p}}({\bf x}_i, {\bf p}_i,t_i)\nonumber\\
&&\times\int \Pi_{j=1}^Np_j^\mu d^3\sigma_{j\mu}\frac{d^3{\bf p}_j}{E_j}f_{n/\bar{n}}({\bf x}_j, {\bf p}_j,t_j)\nonumber\\
&&\times f_A({\bf x}_1^\prime, ... ,{\bf x}_Z^\prime,{\bf x}_1^\prime, ... ,{\bf x}_N^\prime; {\bf p}_1^\prime, ... ,{\bf p}_Z^\prime,{\bf p}_1^\prime, ... ,{\bf p}_N^\prime;t^\prime)\nonumber\\
&&\times \delta^{(3)}\left({\bf P}_A-\sum_{i=1}^Z{\bf p}_i-\sum_{j=1}^N{\bf p}_j\right),
\end{eqnarray}
where $g_{A}=(2J_{A}+1)/2^{A}$ is the statistical factor for $A$ spin 1/2
nucleons to form a nucleus of angular momentum $J_{A}$. The
$p^{\mu }d^{3}\sigma _{\mu }$ denotes the differential hyper-surface of emitted
nucleons such that
$\int p^{\mu }d^{3}\sigma _{\mu }\frac{d^{3}{\mathbf{p}}}{(2\pi )^{3}E}f_{p,
\bar{p}}({\mathbf{x}}, {\mathbf{p}},t)=N_{p, \bar{p}}$ and
$\int p^{\mu }d^{3}\sigma _{\mu }\frac{d^{3}{\mathbf{p}}}{(2\pi )^{3}E}f_{n}({
\mathbf{x}},\allowbreak  {\mathbf{p}},t)=N_{n,\bar{n}}$. The ${\mathbf{x}}_{i}$ and
${\mathbf{p}}_{i}$ are the coordinate and momentum of $i$-th nucleon in the
frame of the nucleon emission source, while ${\mathbf{x}}_{i}^{\prime }$ and
${\mathbf{p}}_{i}^{\prime }$ are those in the rest frame of the produced nucleus,
which can be obtained from the coordinate ${\mathbf{x}}_{i}$ and momentum
${\mathbf{p}}_{i}$ by the Lorentz transformation. For the proton (neutron) phase
space distributions $f_{p/n}({\mathbf{x}},{\mathbf{p}},t)$, they are obtained in
the present study from the positions, momenta and times of protons and
neutrons at their last scatterings, i.e., their kinetic freeze out, in
the UrQMD or AMPT model.

Following Ref.~\cite{Chen:2003qj}, we take the Wigner functions of deuteron and triton to be Gaussian and use the same proton and neutron masses. For the deuteron, its Wigner function is then 
\begin{eqnarray}
f_2(\boldsymbol\rho,{\bf p}_\rho)=8\exp\left[-\frac{\boldsymbol\rho^2}{\sigma_{d}^2}-{\bf p}_\rho^2\sigma_{d}^2\right],
\label{two}
\end{eqnarray}
with the relative coordinate $\boldsymbol\rho$ and the relative momentum ${\bf p}_\rho$ defined as
\begin{eqnarray}
\boldsymbol\rho=\frac{1}{\sqrt{2}}({\bf x}_1^\prime-{\bf x}_2^\prime),\quad{\bf p}_\rho=\frac{1}{\sqrt{2}}({\bf p}_1^\prime-{\bf p}_2^\prime).
\end{eqnarray}
For the triton, its Wigner function is
\begin{eqnarray}
&&f_3(\boldsymbol\rho,\boldsymbol\lambda,{\bf p}_\rho,{\bf p}_\lambda)\nonumber\\
&&=8^2\exp\left[-\frac{\boldsymbol\rho^2}{\sigma_{t}^2}-\frac{\boldsymbol\lambda^2}{\sigma_{t}^2}-{\bf p}_\rho^2\sigma_{t}^2-{\bf p}_\lambda^2\sigma_{t}^2\right],
\label{three}
\end{eqnarray}
with the additional relative coordinate $\boldsymbol\lambda$ and relative momentum ${\bf p}_\lambda$  defined as
\begin{eqnarray}\label{rel}
&&{\boldsymbol\lambda=\frac{1}{\sqrt{6}}({\bf x}_1^\prime+{\bf x}_2^\prime-2{\bf x}_3^\prime)\quad{\bf p}_\lambda=\frac{1}{\sqrt{6}}({\bf p}_1^\prime+{\bf p}_2^\prime-2{\bf p}_3^\prime).}\nonumber\\
\end{eqnarray}

The width parameter $\sigma _{d}$ in Eq.~(\ref{two}) is related to the
root-mean-square matter radius $r_{d}$ of deuteron by
$\sigma _{d}=\frac{2}{\sqrt{3}}r_{d}$. Similarly, the width parameter
$\sigma _{t}$ in Eq.~(\ref{three}) can be related to the root-mean-square
radius $r_{t}$ of triton by $\sigma _{t}=r_{t}$. 
For the values of $r_{d}$ and $r_{t}$, they are taken to be
1.96 fm and 1.59 fm, respectively, from experimental measurements~\cite{Ropke:2008qk}.
We note that because of the very small yields of deuterons and tritons
in heavy-ion collisions at the RHIC BES energies, protons used in the coalescence
processes have negligible effects on the final proton spectra.

It is worth to point out that there is an ambiguity of a prefactor in Eq.(\ref{coal}) for the three-body coalescence production of triton via $p+n+n\rightarrow t$.  In Refs.~\cite{Mattiello:1996gq,Chen:2003ava,Chen:2003qj}, where nucleons are treated as classical particles,  a prefactor of 1/2 is introduced to avoid double counting the neutrons in the production of tritons. This factor is, however, absent in Ref.~\cite{Csernai:1986qf}. According to Refs.~\cite{Sun:2017xrx,Sun:2020uoj}, without this factor the triton yield in the coalescence model using nucleons from a thermally and chemically equilibrated emission source is identical to that in the thermal model when the triton binding energy is neglected. In this work, we only consider the three-body coalescence  channel  for triton production and  do not include a prefactor of 1/2 in Eq.(\ref{coal}) as in ~Refs.~\cite{Csernai:1986qf,Sun:2017xrx,Sun:2020uoj}.  Compared to our  previous study on triton production from the three-body channel in Ref.~\cite{Zhao:2020irc}, which  includes  the prefactor  of  1/2,  both  the  triton yield and  the  ratio $N_tN_p/N^2_d$ at high multiplicity in this work will  thus  be about  a factor of two  larger.  

\begin{figure*}[hbt]
  \centering \includegraphics[scale=0.86]{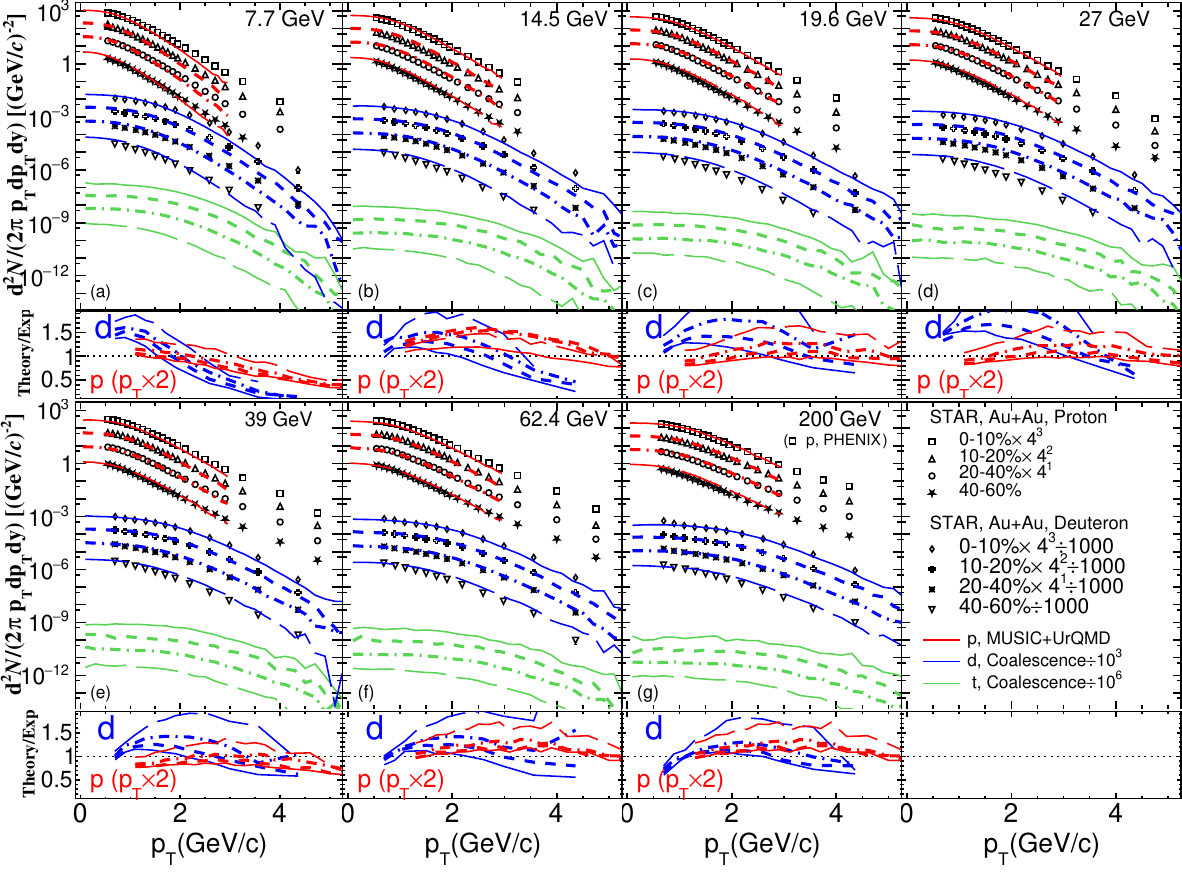}
  \caption{(Color online) Transverse momentum spectra of  protons,  deuterons and tritons in Au + Au collisions at $\sqrt{s_\mathrm{NN}}=$ 7.7, 14.5, 19.6, 27, 39, 62.4, and 200 GeV  from the MUSIC+UrQMD+Coal model (solid lines). Experimental data (symbols)  are taken from the STAR  Collaboration   for deuterons~\cite{Adam:2019wnb}  and   from the STAR and PHENIX Collaborations   for protons~\cite{Adamczyk:2017nof,Adler:2003cb}. The lower panels give the ratio of the
model results to the experimental data. }
  \label{fig:sepctratdp}
\end{figure*}

\subsection{The MUSIC +UrQMD hybrid model for heavy-ion collisions} 
\label{sec:music}

For the description of heavy-ion collision dynamics, we employ the MUSIC+UrQMD
hybrid model~\cite{Shen:2020jwv,Schenke:2010rr,Schenke:2010nt,Paquet:2015lta,Denicol:2018wdp}.
The (3+1)D viscous hydrodynamics in MUSIC, which conserves both the energy-momentum
and baryon number of produced QGP, is developed for describing the collective
dynamics of QGP and the soft hadrons produced from the hadonization hypersurface.
At the RHIC BES energies, this hybrid model uses a smooth initial condition
with the net baryon density taken from an extended 3D Glauber model as the input
for the subsequent hydrodynamic evolution~\cite{Shen:2020jwv}. For the
present study, we use a crossover type of equation of state (EoS) (\texttt{NEOS-BQS})
with the strangeness neutrality condition of vanishing net strangeness
density, $n_{s}=0$, and the net electric charge-to-baryon density ratio
$n_{Q}=0.4n_{B}$~\cite{Monnai:2019hkn}. We use such an EoS without a QCD
critical point because the aim of our study is to provide a reliable baseline
calculation without any critical effects on light nuclei production in
heavy-ion collisions at energies from the RHIC BES program. Following Ref.~\cite{Shen:2020jwv},
we include in the hydrodynamic evolution a temperature and baryon chemical
potential dependent specific shear viscosity $\eta /s$. After the hydrodynamic
evolution, we convert fluid cells to hadrons when their energy densities
drop below the switching energy density $e_{\mathrm{sw}}$=0.26 GeV/$\mathrm{fm^{3}}$.
For the evolution and decoupling of the resulting hadronic matter, the
MUSIC code switches to the microscopic UrQMD hadron cascade model~\cite{Bass:1998ca,Bleicher:1999xi,UrQMD}.
Finally, we obtain the phase-space distributions of kinetic freeze-out
nucleons for the coalescence model calculations.

\begin{figure*}[htb]
  \includegraphics[scale=0.7]{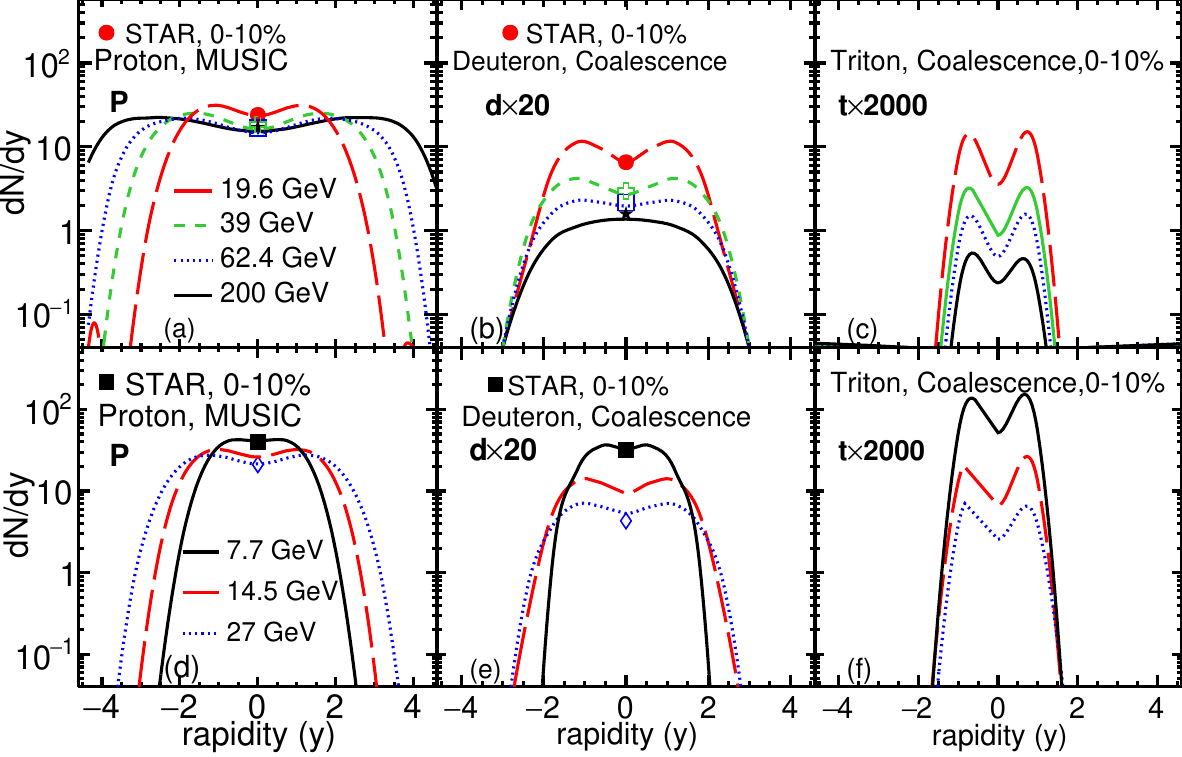}
  \caption{(Color online) The rapidity distribution $dN/dy$ of protons,  deuterons, and   tritons in 0-10\% Au + Au collisions  from the MUSIC+UrQMD+Coal. model (lines).  The experimental data for protons and deuterons are taken from Refs.~\cite{Adamczyk:2017nof,Adler:2003cb,Adam:2019wnb,Zhang:2019wun}.}
  \label{fig:dndy}
\end{figure*}

We would like to emphasize that the MUSIC+UrQMD hybrid model employed in
this paper has achieved a quantitative description of soft particle production,
including the $p_{T}$-spectra and collective flow, in the central and peripheral
Au+Au collisions at $\sqrt{s_{\mathrm{NN}}}= 7.7-200$ GeV, as demonstrated
in Ref.~\cite{Shen:2020jwv}. This indicates that the MUSIC+UrQMD hybrid
model can provide, for the coalescence model calculations, the proper phase-space distributions of nucleons at the kinetic freeze-out, i.e., their last scatterings in the center-of-mass frame of the hadronic matter or the fireball frame.
In the coalescence calculations, the coalescing nucleons
are boosted to their center of mass frame with those of earlier times further
propagated to the time of the latest one. These nucleons then have the probability
to form a light nucleus that is given by the product of their spin statistical factor
to form the nucleus and the Wigner function of the nucleus evaluated at their
equal-time spatial and momentum separation distances. The position and momentum of formed nucleus are finally Lorentz transformed back to the fireball frame.
 
\section{RESULTS}\label{sec:results}
In this section, we study the transverse momentum spectra, the rapidity
distribution $dN/dy$, the mean transverse momentum
$\left <p_{T}\right >$ as well as the multiplicity dependence of the coalescence
parameters $\sqrt[A-1]{B_{A}}$ $(A=2,3)$ and the yield ratio
$N_{t} N_{p}/N_{d}^{2}$ of triton, deuteron and proton in Au+Au collisions
at $\sqrt{s_{\mathrm{NN}}}=$ 7.7--200 GeV. For the yield ratio, we also include
results for Pb+Pb collisions at 2.76 TeV from the coalescence calculations based on the VISHNU hybrid model.

\subsection{Transverse momentum spectra and 
rapidity distributions}

Figure~\ref{fig:sepctratdp} shows the transverse momentum spectra of protons, deuterons and tritons from central to peripheral Au+Au collisions at $\sqrt{s_\mathrm{NN}}=$ 7.7, 14.5, 19.6, 27, 39, 62.4, and 200 GeV.  It is seen that the MUSIC+UrQMD hybrid model can quantitatively describe  measured  proton  spectra  below 2.5 GeV  but  slightly underestimates the data at higher $p_T$, where the quark recombination contribution  has been shown to become increasingly important~\cite{Greco:2003mm,Greco:2003vf,Fries:2003kq,Hwa:2004ng,Fries:2008hs,Zhao:2019ehg,Zhao:2021vmu} \footnote{The proton spectra in both the STAR data and the model calculations have been corrected
by subtracting the weak decay feed-down contributions. The weak decay feed-down correction to the STAR proton spectra reported in Ref.~\cite{Adamczyk:2017nof} is based on the UrQMD+GEANT simulation.}  With the phase-space distributions of protons and neutrons at kinetic freeze-out from the MUSIC+UrQMD hybrid model,  we calculate the spectra of deuterons and tritons within the framework of nucleon coalescence model.  As shown by blue  lines in Fig. 1, without any free parameters, the  coalescence model calculations can reasonably describe the $p_T$-spectra of deuterons measured by the STAR Collaboration at 0-10\%, 10-20\%, 20-40\% and 40-60\% centrality bins in Au+Au collisions at $\sqrt{s_\mathrm{NN}}= 7.7-200$ GeV. The ratio  of the theoretical result to the experimental data, denoted by Theory/Exp in Fig.~\ref{fig:sepctratdp}, is mostly between 0.5 and 1.5. 

Figure~\ref{fig:dndy} shows the rapidity distribution  $dN/dy$  of protons, deuterons, and tritons in 0-10\% Au+Au collisions at  $\sqrt{s_\mathrm{NN}}= 7.7-200$ GeV.  It is seen that our model calculations give an excellent description of the STAR data at mid-rapidity.  For the proton yield at mid-rapidity, it is determined in the (3+1)D MUSIC model by the interplay between  the initial  baryon stopping and  thermal production  at chemical freeze-out.   At  low collision energies, the initial baryon stopping and baryon current evolution in the hydrodynamic phase are more important than   thermal production at particalization, resulting in a larger proton yield as observed in experiments.  The calculated $dN/dy$ of deuterons and tritons also show a similar trend with respect to the change in the collision energy, which  are again consistent with the STAR data. For the predicted shapes of the $dN/dy$ of protons, deuterons and tritons at $\sqrt{s_\mathrm{NN}}$=7.7-200 GeV, they become wider with increasing collision energy, which can be tested in upcoming experimental measurements.

\begin{figure}[thp]
  \includegraphics[scale=0.4]{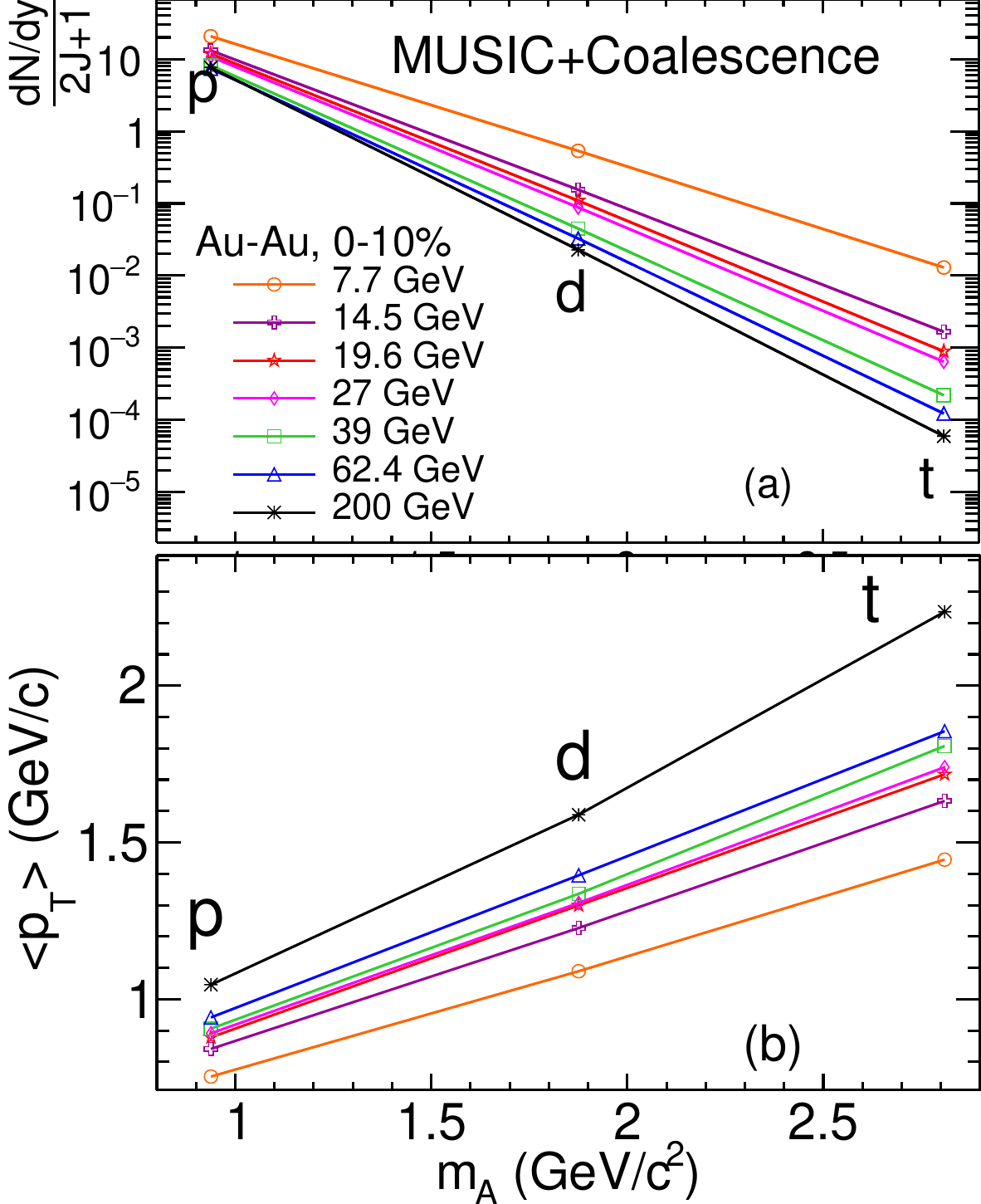}
  \caption{(Color online) The atomic mass dependence of  $dN/dy$ (upper panel) and $\left< p_T\right> $ (lower panel) of protons,  deuterons, and   tritons   in  0-10\% Au + Au collisions  from the MUSIC+UrQMD+Coal model.}
  \label{fig:meanpt}
\end{figure}

In the upper panel (a) of Fig.~\ref{fig:meanpt}, we show the atomic mass ($m_A$, $A=1, 2, 3$ for proton, deuteron and triton) dependence of the rapidity distribution per degree of freedom, $\frac{dN/dy}{2J+1}$, in 0-10\% Au+Au collisions at $\sqrt{s_{NN}}$=7.7-200 GeV.  An  exponentially  decreasing  $\frac{dN/dy}{2J+1}$ with the atomic  mass is observed  as expected from  the nucleon coalescence model, in which the binding energies of produced light nuclei  are ignored. The lower panel (b) shows the  dependence of the  mean transverse momentum $\left<p_T\right>$  on  the atomic  mass $m_A$.  The $\left<p_T\right>$ increases with increasing $m_A$ because  light nuclei are produced in the coalescence model from nucleons close in phase space, thus enhancing their momenta. Besides, larger values of $\left<p_T\right>$ are observed at higher collision energies due to the stronger radial flow, which pushes particles  to larger $p_T$. 

\subsection{Coalescence parameters} 

In interpreting light nuclei production from nuclear reactions, one usually expresses the yield $d^3N_A/d^3\bf p_A$ of a nucleus with the mass number $A=Z+N$ and momentum $ \bf p_A$ in terms of the yield $d^3N_{\rm p}/d^3\bf p_{\rm p}$  of protons with momentum $\bf p_{\rm p}$ and the yield $d^3N_{\rm n}/{d^3\bf p_{\rm n}}$  of neutrons with momentum $\bf p_{\rm n}$ in terms of the coalescence parameter $B_A$ as follows: 
\begin{eqnarray}
E_{A}\frac{\mathrm{d}^{3}N_{A}}{\mathrm{d^3}{\bf p}_{A}}&=&B_{A}{\left(E_{\mathrm{p}}\frac{\mathrm{d}^{3}N_{\mathrm{p}}}{\mathrm{d^3}{\bf p}_{\mathrm{p}}}\right)^{Z}}{\left(E_{\mathrm{n}}\frac{\mathrm{d}^{3}N_{\mathrm{n}}}{\mathrm{d^3}{\bf p}_{\mathrm{n}}}\right)^{A-Z}}\nonumber\\
&\approx&B_{A}{\left(E_{\mathrm{p}}\frac{\mathrm{d}^{3}N_{\mathrm{p}}}{\mathrm{d^3}{\bf p}_{\mathrm{p}}}\right)^{A}}\left\vert_{\bf{p}_{\mathrm{p}}=\bf{p}_{\mathrm{n}}=\frac{\bf{p}_{A}}{A}} \right. ,
\label{eq:BA}
\end{eqnarray}
where  $E_{p,n}$ are  the proton and neutron energies. The coalescence parameter $B_A$ (A=2,3) contains information on the freeze-out properties of the nucleon emission source, i.e., ${B_A}\propto V_{\text{eff}}^{1-A}$ with $V_{\text{eff}}$ being  the effective volume of the nucleon emission source~\cite{Csernai:1986qf,Scheibl:1998tk,Bellini:2018epz,Wang:2020zaw}.

\begin{figure}[htb]
  \includegraphics[scale=0.4]{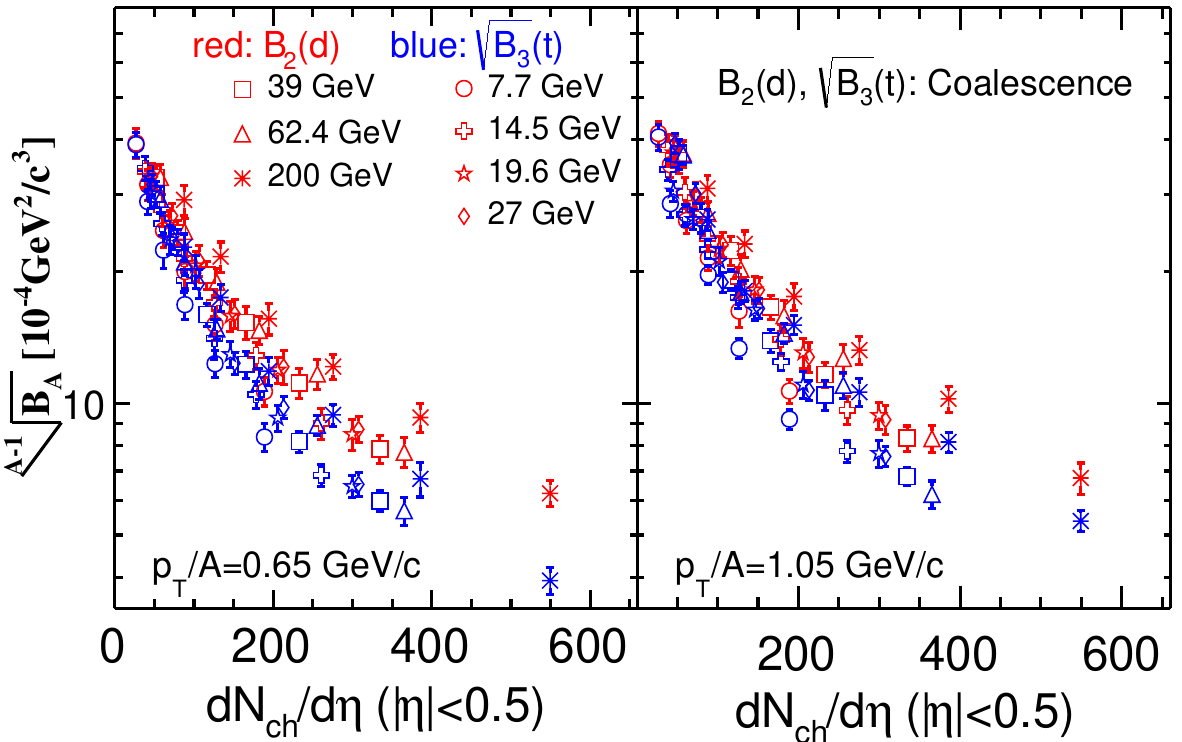}
  \caption{(Color online) Multiplicity dependence of the coalescence parameters $B_2(d)$ and $\sqrt{B_3(t)}$ at $p_T/A=$ 0.65 GeV/c and  $p_T/A=$ 1.05 GeV/c in  Au+Au collisions from 7.7 GeV to 200 GeV, calculated by the coalescence model. }
  \label{fig:bn}
\end{figure}

Figure~\ref{fig:bn} shows  the multiplicity  dependence of the  coalescence parameters $B_2(d)$ and $\sqrt{B_3(t)}$ at $p_T/A$=0.65 GeV/c (left) and  $p_T/A$=1.05 GeV/c (right) in Au+Au collisions from 0-10\% to 50-60\% centrality bins, with $A=2$ for deuterons and $A=3$ for tritons. The $B_2(d)$  and $\sqrt{B_3(t)}$ from our model calculations decrease monotonically with  increasing multiplicity, and this is because the size of the nucleon emission  source increases monotonically with increasing  multiplicity. It is seen that the $\sqrt{B_3(t)}$ decreases  more rapidly than $B_2(d)$ as a function of multiplicity. Specifically, the values of $\sqrt{B_3(t)}$ are very close to $B_2(d)$ at low multiplicities, but they start to deviate when $dN_{ch}/d\eta>200$, which can be attributed to the different deuteron and triton sizes~\cite{Wang:2020zaw}. This is also consistent with the multiplicity dependence of the yield ratio $N_t N_p/N_d^2$ shown in the next subsection. The slightly smaller coalescence parameters at $p_T/A$=0.65 GeV/c  than those at $p_T/A$=1.05 GeV/c are consistent with the decreasing correlation lengths (HBT radii)   with mean pair transverse momentum extracted in the STAR expeiments~\cite{Adamczyk:2014mxp}.  For a fixed multiplicity, the coalescence parameter is systematically larger at a higher collision energy. For example, at $dN_{ch}/d\eta\sim250$, the  $B_2(d)$ in collisions at $\sqrt{s_\mathrm{NN}}$=200 GeV is clearly greater than that at  $\sqrt{s_\mathrm{NN}}$=14.5 GeV. This is attributed to the stronger  collective expansion at the higher collision energy, which leads to a decrease of the effective volume at kinetic freeze-out because of the focusing effect due to the larger transverse flow, although the total volume is larger at higher collision energy~\cite{Adamczyk:2014mxp}. 

Although the coalescence parameters $B_2(d)$ and $\sqrt{B_3(t)}$ are expected to depend inversely on the effective volume of the nucleon emission source, their dependence on the charged particle multiplicity, which is proportional to the true volume of the emission source, is nontrivial because of the complicated relation between the effective and true volumes of the emission source. For the coalescence parameters $B_2(d)$ and $\sqrt{B_3(t)}$ from the MUSIC+UrQMD hybrid model shown in Fig.~\ref{fig:bn}, it is found that their scaled values $B_2(d)(dN_{\rm ch}/d\eta)^{0.765}$ and $\sqrt{B_3(t)}(dN_{\rm ch}/d\eta)^{0.765}$ then depend only weakly on the charged particle multiplicity.

\subsection{The yield ratio $N_t  N_p/N_d^2$}


\begin{figure*}[th]
\center
  \includegraphics[scale=0.65]{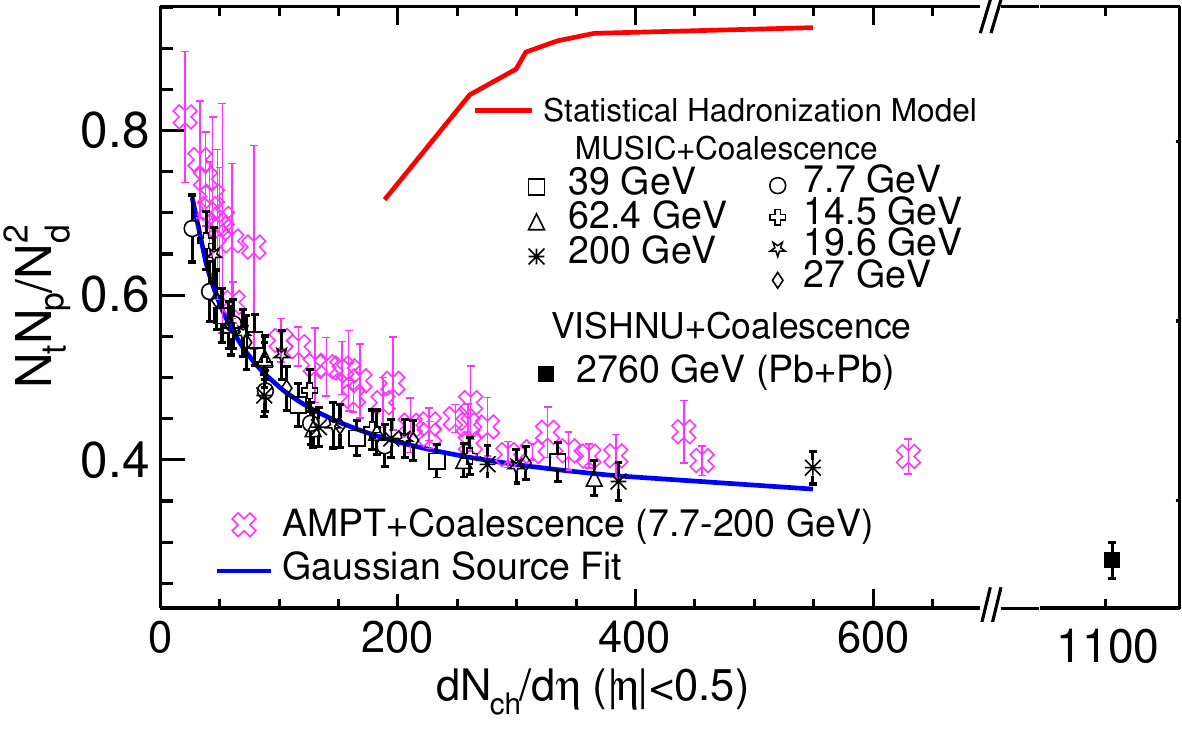}
  \caption   {(Color online)  Multiplicity dependence of the yield ratio $N_t  N_p/N_d^2$ in  Au+Au collisions at $\sqrt{s_\mathrm{NN}}=$ 7.7 - 200 GeV and Pb+Pb collisions at $\sqrt{s_\mathrm{NN}}=$ 2.76 TeV calculated from the nucleon coalescence model using kinetically freeze-out nucleons from the MUSIC+UrQMD, VISHNU and AMPT models, as well as the statistical model that includes only stable nuclei~\cite{Vovchenko:2020dmv}.  Also shown by the blue solid line is the result from fitting the results of MUSIC+UrQMD hybrid model by Eq.~(\ref{eq:doubleratio}) with  the mutiplication factor $p_0=0.683$ and using the relation $R=0.547(dN_{\rm ch}/d\eta)^{0.331}$ fm between the radius $R$ and the charged particle multiplicity $dN_{\rm ch}/d\eta$.}
  \label{fig:ratio}
\end{figure*}

We have also studied the charged particle multiplicity dependence of the yield ratio $N_t  N_p/N_d^2$ in Au+Au at $\sqrt{s_\mathrm{NN}}$=7.7-200 GeV from the MUSIC+UrQMD hybrid model, and it is shown in  Fig.~\ref{fig:ratio}   by black open symbols.
The yield ratio $N_t  N_p/N_d^2$ is seen to decrease monotonically with increasing charged particle multiplicity, and the same behavior is seen for different collision centralities and energies as they almost lie on the same curve.  This scaling behavior in the multiplicity dependence indicates that the  yield  ratio $N_t  N_p/N_d^2$  in relativistic heavy-ion collisions  is insensitive to the baryon density and   collective flow effects.

The above scaling behavior can be qualitatively understood by considering a  thermally equilibrated spherical Gaussian nucleon emission source with a width parameter or radius  $R$ at certain temperature and nucleon chemical potentials, i.e., $\propto e^{-r^2/(2R^2)}$.  The yield ratio $N_t  N_p/N_d^2$ in the coalescence model can then be obtained analytically~\cite{Sun:2018mqq}, and it is given by 
\begin{eqnarray}
 \frac{N_tN_p}{N_d^2}=\frac{4}{9}\left(\frac{1+\frac{2r_d^2}{3R^2}}{1+\frac{r_t^2}{2R^2}}\right)^3=\frac{4}{9}\left(1+\frac{\frac{4}{3}r_d^2-r_t^2}{2R^2+r_t^2}\right)^3, 
\label{eq:doubleratio}
\end{eqnarray}
Since  $(2/\sqrt{3})r_d=2.26$  fm is greater than  $r_t=1.59$  fm,  the yield ratio thus 
decreases with increasing source size or the particle multiplicity in heavy-ion collisions as 
one expects $R\propto(dN_{\rm ch}/d\eta)^{1/3}$ for a static source.   Eq.
(\ref{eq:doubleratio}) also shows that this behavior is independent of the temperature and 
nucleon chemical potentials of the emission source. This result is similar to the multiplicity 
scaling behavior of the yield ratio $ N_tN_p/N_d^2$ seen in the coalescence model 
calculations based on kinetic freeze-out nucleons from the MUSIC+UrQMD hybrid model.   
Eq.~(\ref{eq:doubleratio}) further shows that the ratio $ N_tN_p/N_d^2$ has an 
asymptotic value of 4/9 when $R\gg r_d,r_t$, which is again similar to the value
shown in Fig.~\ref{fig:ratio}.  However, fitting the ratio $N_tN_p/N_d^2$ according to Eq.(\ref{eq:doubleratio}) using the relation $R= \alpha ( dN_{\rm ch}/d\eta)^{ \beta} $ fm  by treating $\alpha$ and $\beta$ as parameters leads to an unrealistic large radius for a given charged particle multiplicity.  This indicates the inadequacy of using a static Gaussian form to describe the emission source from the hybrid and AMPT models, which includes dynamic effects such as the space and momentum correlations. As a possible improvement to the modeling of the emission source, we multiply Eq.(\ref{eq:doubleratio}) by a factor $p_0$.  A good description of the MUSIC+UrQMD hybrid model results in Fig.~\ref{fig:ratio} can then be obtained with $p_0=0.683$ and a more realistic relation of $R=0.547(dN_{ch}/d\eta)^{0.331}$ as shown by the blue solid line in the figure.  We note that if we had taken the nucleon emission source to be uniform in space as assumed in Refs.~\cite{Sun:2017xrx,Sun:2018jhg}, the yield ratio $N_tN_p/N_d^2$ would have a smaller asymptotic value of $(4/9)\times(3/4)^{3/2}=\frac{1}{2\sqrt{3}}$.  Therefore, including the surface diffuseness of the nucleon emission source, which is expected in realistic heavy-ion collisions, enhances this value. Also, the size effects discussed here will be even stronger for hypernuclei due to their larger sizes~\cite{Sun:2018mqq}.

Similar results are obtained from the nucleon coalescence model using kinetic freeze-out nucleons from the AMPT model as shown by pink open symbols in Fig.~\ref{fig:ratio}. The AMPT model is a multiphase transport model that uses initial conditions from the HIJING model~\cite{Gyulassy:1994ew,Wang:1991hta}, the ZPC model ~\cite{Zhang:1997ej} for the partonic cascade, and the ART model~\cite{Li:1995pra} for the hadronic transport as well as a  spatial quark coalescence model to convert kinetic freeze-out quarks and anti-quarks to  initial hadrons.

Also shown by the filled square in Fig.~\ref{fig:ratio} is the yield ratio $N_t  N_p/N_d^2$ for Pb+Pb at $\sqrt{s_\mathrm{NN}}$=2.76 TeV calculated from the nucleon coalescence model based on kinetic freeze-out nucleons from the 2D VISHNU hybrid model~\cite{Zhao:2017yhj,Zhao:2018lyf}.  Its small value at a large charged particle multiplicity follows nicely the scaling behavior seen in the results from the MUSIC+UrQMD hybrid model at lower multiplicities.  The VISHNU model is  an  event-by-event hybrid model that combines 2+1D viscous hydrodynamics VISH2+1 for the QGP expansion  with the UrQMD model  for  the evolution of  the hadronic   matter. In the hydrodynamic part  of VISHNU, it uses  the crossover type equation of state s95-PCE~\cite{Huovinen:2009yb} and neglects the net  baryon  density  and  heat  conductivity effects in heavy-ion collisions  at  the  LHC energies. The above yield ratio $N_t  N_p/N_d^2$  from the VISHNU + coalescence model for central Pb+Pb collisions at $\sqrt{s_\mathrm{NN}}=2.76$ TeV is close to the value at central Au+Au collisions at 200 GeV in the present study using the MUSIC+coalescence model. However, the recent ALICE measurement of the $N_{^3He}  N_p/N_d^2$ ratio in central Pb+Pb at $\sqrt{s_\mathrm{NN}}$=2.76 TeV reaches a value around 0.9~\cite{Adam:2015vda}, which is much larger than the  theoretical value of about 0.33 from Refs.~\cite{Zhao:2017yhj,Zhao:2018lyf}.  Although the uncertainty in the ALICE data is still large, the  almost factor of three difference between the coalescence model calculation and  the  experimental data at LHC needs to be understood in the future. 

We note that  the thermal model gives an opposite behavior in  the multiplicity dependence of $N_t  N_p/N_d^2$, i.e., it increases with increasing multiplicity  as shown by the red solid line in Fig.~\ref{fig:ratio}~\cite{Vovchenko:2018fiy,Vovchenko:2020dmv}.  According to Eq.(\ref{eq:doubleratio}), the larger $N_tN_p/N_d^2$ at lower multiplicity is due to the larger deuteron than triton radius. This enhancement in the value of $N_tN_p/N_d^2$ would disappear if the deuteron and triton radii were similar. Indeed, changing the triton radius to that of deuteron in the coalescence model calculations using nucleons from the AMPT model, we have obtained an almost constant value of about 0.33 for all particle multiplicities.  This result strongly indicates the importance of the size effect on coalescence production of light nuclei in heavy-ion collisions. The upcoming measurements of   $N_t  N_p/N_d^2$  as a function of  multiplicity would thus help discriminate different production mechanisms of light nuclei in heavy-ion collisions. We would like to point out that neither the hydrodynamic model  nor  the  AMPT model  used  in our study includes any dynamical density fluctuations related to the critical point or first-order phase transition of the QCD matter. Therefore, our results can serve as the baseline predictions for the yield ratio of light nuclei in the search for the possible QCD critical point from experimental beam energy scan of heavy-ion collisions.

\section{Summary}\label{sec:summary}

In this paper, we have studied the multiplicity  dependence  of light nuclei production 
in Au+Au collisions at $\sqrt{s_\mathrm{NN}}=$ 7.7 - 200 GeV  by using the nucleon coalescence model. With the phase-space distributions of protons and neutrons at kinetic freeze-out taken from the MUSIC+UrQMD model with a smooth crossover transition from the QGP to hadronic matter, the theoretical  results nicely reproduce the measured  $p_T$-spectra of protons and deuterons in 0-10\%, 10-20\%, 20-40\% and 40-60\% Au+Au collisions at $\sqrt{s_\mathrm{NN}}= 7.7-200$ GeV. The coalescence parameters $B_2(d)$ and $\sqrt{B_3}(t)$ for deuteron and triton production, respectively,  are found to decrease monotonically with increasing multiplicity. Using also kinetic freeze-out nucleons from the VISHNU hybrid model with a crossover transition for the coalescence production of deuterons and tritons in Pb+Pb collision at $\sqrt{s_\mathrm{NN}}=2.76$ TeV, the yield ratio $N_tN_p/N^2_d$ in heavy-ion collisions  from 7.7 GeV to 2.76 TeV is found to exhibit a scaling behavior  in its multiplicity dependence, i.e., decreasing monotonically with increasing charged-particle multiplicity,  but not much on  the collision energy and centrality. This scaling behavior is further seen in calculations that use the kinetic freeze-out nucleons from the AMPT model for deuteron and triton production in heavy-ion collisions at RHIC energies.  The predicted multiplicity scaling of the yield ratio $N_tN_p/N^2_d$ can be used to validate the production mechanism of light nuclei and to extract their sizes in relativistic heavy-ion collisions.   As to the collision energy dependence of this yield ratio predicted by the MUSIC+UrQMD hybrid model and the AMPT model with a crossover QGP to hadronic matter transition as used in the present study, both give essentially a constant value~\cite{Zhao:2020irc,Sun:2020uoj}.  Although these models cannot describe the non-monotonic collision energy dependence in the preliminary STAR data~\cite{Zhang:2020ewj}, which has been argued to be related to dynamical density fluctuations  from the critical point and first-order phase transitions of QCD~\cite{Sun:2017xrx,Sun:2018jhg,Sun:2020zxy},  results  from these models  can serve as a baseline against which one can compare  the experimental data from the beam energy scan of heavy-ion collisions to search for the critical point in the QCD phase diagram. 

\section*{Acknowledgements}
We thank Chun Shen for helpful discussions. This work is supported in part by the National Key Research and Development Program of China (Grant No.  2020YFE0202002 and 2018YFE0205201), the National Natural Science Foundation of China (Grant No. 11890711, 11861131009, 11935007, 11221504 and 11890714), US DOE under grant No. DE-AC02-05CH11231, US NSF under grant Nos. ACI-1550300 and OAC-2004571, the UCB-CCNU Collaboration Grant. K. J. S. and C. M. K. were  supported  in  part  by  the  US  Department of Energy under  Award No. DE-SC0015266  and the Welch Foundation under Grant No. A-1358.

\bibliography{bibliography}

\end{document}